%% file: photon.tex
RU98-5-B
\input gb.tex
\finalversion
%
%
\input epsf
%
%
%
%
\begintitle
On T-Violation Signals Originating
from Magnetic Orders through 
Double Exchange Mechanism
\endtitle

\beginauthor
H. C. Ren
\endauthor

\beginaddress
Department of Physics, The Rockefeller University,
New York, NY 10021, USA
\endaddress

\beginauthor
M. K. Wu
\endauthor

\beginaddress
Department of Physics, National Tsing Hua University,
Hsinchu, Taiwan, ROC
\endaddress

\beginabstract

The photon self energy tensor in a crystal with double exchange and 
ferromagnetic or canting magnetic orders is analyzed at zero temperature 
with emphasis on the signal of the time-reversal invariance breaking.
Three comparable contributions, spin susceptibility, local magnetic 
field and spin-orbit coupling are estimated and the prospect of their
experimental detection is discussed.

\endabstract

\section{1. Introduction}

Recently, there are growing interests in the double exchange mechanism 
proposed by Zener [1], Anderson and Hasegawa [2] over forty years ago. It 
stems from the possible relevance of the mechanism to the colossal 
magneto-resistance of the single perovskite material $La_{2-x}Ca_xMnO_3$ 
[3] and is conjectured also to be responsible to the newly synthesised 
double perovskite material $(Sr, Ba)_2YRu_{1-x}Cu_xO_6$ [4]. The physical 
picture of the double exchange mechanism in the latter 
system lies in the outer electronic configuration of the $Ru$-ion. Prior 
to doping, the $Ru$-ion carries a charge +5 and its outer electronic 
configuration is $4d^3$, the spin of each electron are forced parallel by 
strong Hund's rule coupling and the three electrons fills in a closed 
multiplet of the cubic group. The total spin 
of each $Ru^{+5}$ is 3/2 and interacts with each other through an 
antiferromagnetic super-exchange. The parent compound is therefore an 
antiferromagnetic insulator. When the charge 
balance is offset by doping of lower valence transition elements, $Cu$, 
holes are introduced to the outer electronic configuration and they 
start to hop. The system becomes a conductor and may even be a 
superconductor below certain temperature. The spins 
of these holes couple strongly with the $Ru^{+5}$ ions via Hund's rule
and bring in ferromagnetic couplings between ion spins. The prototype 
Hamiltonian of this mechanism reads

$$H=\sum_{ij}J_{ij}\vec S_i\cdot\vec S_j-\mu_c\sum_j\vec B_j\cdot\vec S_j$$
$$-{1\over 2}\sum_{ij}t_{ij}(\Psi_i^\dagger \Psi_j+\Psi_j^\dagger \Psi_i)
+{1\over 2}\lambda\sum_j\vec S_j \Psi_j^\dagger\vec\sigma \Psi_j
+H_{Coul.}. \eqno(1.1)$$ 
where $\vec S_j$ is the total spin operator of the ion core, $\Psi_j$,
$\Psi_j^\dagger$ are the creation and the annihilation operators of 
the itinerant carriers, each of two components with spin up and spin 
down, $J_{ij}$ denotes the antiferromagnetic exchange, $t_{ij}$ denotes
the hopping amplitude of the itinerant electrons and $\lambda$ denotes 
the Hund's rule coupling of the itinerant electrons with the ion cores.
The Coulomb interaction, $H_{Coul.}$, is given by 
$$H_{Coul.}={e^2\over 8\pi\varepsilon}\Big(U\sum_{j,\sigma,\sigma^\prime}
:\Psi_{j\sigma}^\dagger\Psi_{j\sigma}
\Psi_{j\sigma^\prime}^\dagger\Psi_{j\sigma^\prime}:+
\sum_{i\neq j,\sigma,\sigma^\prime} {:\Psi_{i\sigma}^\dagger\Psi_{i\sigma}
\Psi_{j\sigma^\prime}^\dagger\Psi_{j\sigma^\prime}:\over 
|\vec R_i-\vec R_j|}\Big),\eqno(1.2)$$  with the first term the 
on-site Coulomb repulsion and $\varepsilon$ the dielectric constant produced 
by the ion cores. 

The competition between the antiferromagnetic 
exchange and the ferromagnetic coupling may lead to a canting magnetic 
order below certain temperature, as was analyzed by de-Gennes [5] and 
supported by specific heat measurement of the double perovskite materials
[6]. In the previous paper [7], hereby referred to as paper I, we studied the 
magnetic ordering of the Hamiltonian (1.1). Spinwave spectrum was 
calculated from the Hamiltonian (1.1) with a canting magnetic order. 
The important role played by long range Coulomb interaction in 
maintaining such an order is emphasized. We also performed extensive 
Monte Carlo simulations on de-Gennes's effective magnetic Hamiltonian
and computed the specific heat as a function of temperature. All our 
results were found consistent with the experimental measurements 
of the specific heat and the magnetic susceptibility. 

In this paper, we shall address to the optical properties of the unusual 
magnetic ordering caused by the double exchange. The characteristic 
term of the Hamiltonian (1.1) is the Hund's rule coupling term. A 
reasonable estimate [2] gives rise to $\lambda S\sim 1$eV. In the paramagnetic 
phase, such a term causes strong scattering of itinerant holes. In the 
ordered phase with ferromagnetic order or canting, such a term represents 
an effective homogeneous magnetic field acting on the hole spins only. With 
the magnetic moment of a free electron, $2\mu_{{\rm{Bohr}}}$, the 
equivalent magnetic field strength is as high as 
$$B_{{\rm{equiv}}}\sim 10^4{\rm{Tesla}},\eqno(1.3)$$ 
($2\mu_{{\rm{Bohr}}}B_{{\rm{equiv}}}\sim \lambda S$). 
Detectable effect signaling the violation of time reversal invariance 
is expected and such effects are the main theme of this work.

In the following two sections, we shall couple the system described by 
the model Hamiltonian (1.1) with radiation field calculate the photon 
self energy function of the medium. A f.c.c cubic structure 
like the double perovskite materials and and a tetragonal layer type 
magnetic ordering are assumed for concreteness. In contrast 
to the ordinary magnetooptical effect, the equivalent magnetic field
(1.3), essentially rooted in the Coulomb forces, does not acts on 
the orbital motion directly. The optical activity comes from three 
different sources: 1) the Pauli term; 2) the secondary magnetooptical 
effect because of the real magnetic field produced by ordered spins; 3) 
a spin-orbital coupling which comes together with the Pauli-term in 
the nonrelativistic approximation of the Dirac electrons. The prospect 
of the experimental detection, the relevance of possible genuine 
superconductivity order and the validity of various approximation 
we made are discussed in the final section. 

\section{2. The Electromagnetic Coupling and Its Parabolic Approximation}

The Hamiltonian combining the antiferromagnetic exchange, 
the double exchange and the electromagnetic coupling reads
$$H=\int d^3\vec r{1\over 2}(\vec E_{{\rm{tr.}}}^2+\vec B^2)
+\sum_{ij}J_{ij}\vec S_i\cdot\vec S_j-\mu_i\sum_j\vec B_j\cdot\vec S_j$$
$$+{1\over 2}\sum_{ij}t_{ij}(\Psi_i^\dagger U_{ij}\Psi_j
+\Psi_j^\dagger U_{ji}\Psi_i)
-{1\over 2}\lambda\sum_j\vec S_j \Psi_j^\dagger\vec\sigma \Psi_j
-\mu_h\sum_j\Psi_j^\dagger\vec\sigma\cdot\vec B_j\Psi_j+H_{Coul.},\eqno(2.1)$$
The Coulomb gauge is adapted with $\vec\nabla\cdot\vec A=0$. The transverse
electric field $\vec E_{{\rm{tr.}}}=-{\partial\vec A\over \partial t}$
and the magnetic field $\vec B=\vec\nabla\times\vec A$ where 
$\vec B_j=\vec B(\vec R_j)$ with $\vec R_j$ the position vector of the
$j$th site. The lattice gauge coupling to the carrier field is given by 
$$U_{ij}=\exp\Big[-ie\int_0^1ds\vec l_{ij}\cdot 
\vec A(\vec R_i+s\vec l_{ij})\Big]\eqno(2.2)$$
where $\vec l_{ij}$ denotes the displacement vector from the $i$th site to 
the $j$th site. In what follows, we shall restrict the hopping amplitude
within the nearest bond only.

Following the conventions in I, we introduce the lattice frame 
($x$, $y$, $z$) with $\hat x$, $\hat y$ and $\hat z$ parallel to 
the (100), (010) and (001) directions of the crystal respectively. The 
locations of magnetic ions (total number $={\cal N}$) are given by 
$$\vec R=n_1\vec e_1+n_2\vec e_2+n_3\vec e_3,
\eqno(2.3)$$ where $n_1$, $n_2$ and $n_3$ are integers,
$$\vec e_1={l\over 2}(\hat y+\hat z),\eqno(2.4)$$ 
$$\vec e_2={l\over 2}(\hat z+\hat x),\eqno(2.5)$$ 
$$\vec e_3={l\over 2}(\hat x+\hat y),\eqno(2.6)$$ 
with $l$ the side length of a basic cube ($l=8.16\AA$, for 
$Sr_2YRu_{1-x}Cu_xO_6$). Adapting the tetragonal layer magnetic ordering, 
we divide the lattice 
into two sublattices shown in Fig 1. with $n_1+n_2={\rm{even}}$ on the 
sublattice $A$ and $n_1+n_2={\rm{odd}}$ on the sublattice $B$. 
The expectation value of the ion spins of the sublattice $A$ is 
denoted by $\vec S_A=S\vec\zeta_A$ and that of the sublattice $B$ 
by $\vec S_B=S\vec\zeta_B$. We choose the $\zeta$-axis along the 
direction of $\vec S_A+\vec S_B$ and the $\eta$-axis as 
$$\hat\eta={\vec \zeta_A\times\vec \zeta_B\over\sin\Theta}
\eqno(2.7)$$ with $\Theta$ the mutual angle between $\vec S_A$ and 
$\vec S_B$. We introduce further
$$\hat\xi=\hat\eta\times\hat\zeta,\eqno(2.8)$$
The set ($\hat\xi, \hat\eta, \hat\zeta$) forms a right hand 
coordinate system and will be referred to as the spin frame. 
Without the magnetic anisotropic energy, the orientation 
of the spin frames does not couple with the lattice frame.

\topinsert
\hbox to\hsize{\hss
	\epsfxsize=4.0truein\epsffile{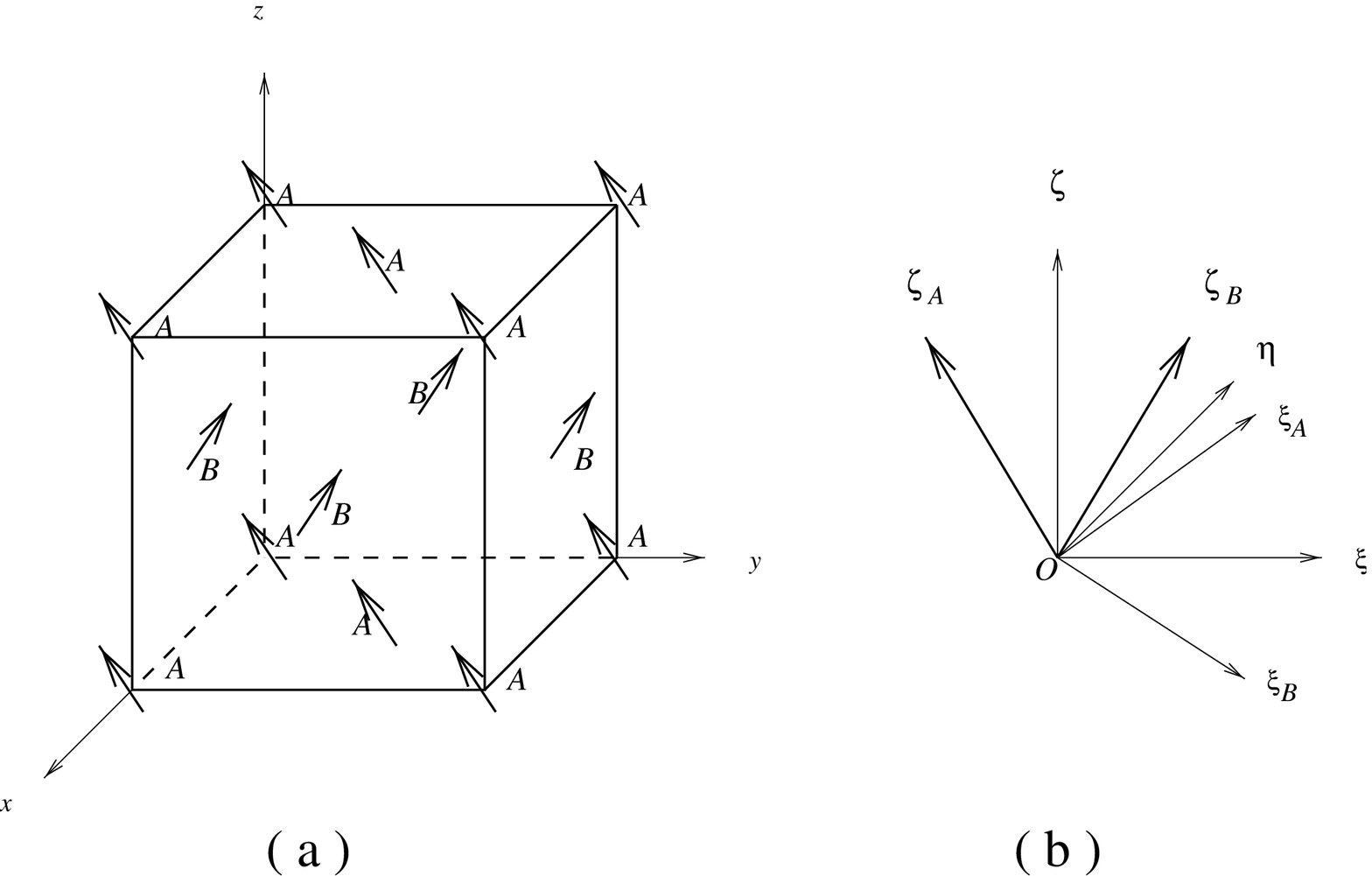}
	\hss}
\begincaption{Figure 1}
(a) The magnetic ordering. The magnetic sites of the sublattices $A$ and 
$B$ are marked explicitly. (b) The spin frame. The angle between $\zeta_A$,
and $\zeta_B$ axes is $\Theta$. 
\endcaption
\endinsert

With large spin approximation, the Hamiltonian (1.1) can be written 
as $$H=H_{{\rm{e.m.}}}+H_{{\rm{s.w.}}}+H_{{\rm{Coul.}}},\eqno(2.9)$$ 
where $H_{{\rm{e.m.}}}$ 
is the part of the Hamiltonian we are concerned with. Its is given by
$$H_{{\rm{e.m.}}}=\int d^3\vec r{1\over 2}(\vec E_{{\rm{tr.}}}^2
+\vec B^2)-{1\over 2}t\sum_{<ab>}(\Psi_a^\dagger U_{ab}\Psi_b
+\Psi_b^\dagger U_{ba}\Psi_a)$$ $$-{1\over 2}t^\prime\Big[\sum_{<aa^\prime>}
(\Psi_a^\dagger U_{aa^\prime}\Psi_{a^\prime}+\Psi_{a^\prime}^\dagger 
U_{a^\prime a}\Psi_a)+\sum_{<bb^\prime>}(\Psi_b^\dagger 
U_{bb^\prime}\Psi_{b^\prime}+\Psi_{b^\prime}^\dagger U_{b^\prime b}
\Psi_b)\Big]$$ $$-{1\over 2}\lambda S(\sum_a\zeta_A \Psi_a^\dagger\vec\sigma 
\Psi_a+\sum_b\zeta_B \Psi_b^\dagger\vec\sigma \Psi_b
-\mu_h(\sum_a\Psi_a^\dagger\vec\sigma\cdot\vec B_a\Psi_a
+\sum_b\Psi_b^\dagger\vec\sigma\cdot\vec B_b\Psi_b),\eqno(2.10)$$
where $t$ and $t^\prime$ denote the interlayer and intralayer hoppings 
respectively. The terms containing spin wave operators are included in 
$H_{{\rm{s.w.}}}$ and have been dealt with in I.

Within Heisenberg representation, the time reversal operator $T$ transforms
$$T\vec A(\vec r,t)T^{-1}=-\vec A(\vec r,-t),\eqno(2.11)$$
$$T\Psi_j(t)T^{-1}=\kappa\sigma_2\Psi_j(-t)\eqno(2.12)$$ and
$$T\Psi_j^\dagger(t)T^{-1}=\kappa^*\Psi_j^\dagger(-t)\sigma_2\eqno(2.13)$$
with $\kappa$ a phase factor. It is easy to see that
$$TH_{{\rm{e.m.}}}T^{-1}\neq H_{{\rm{e.m.}}},\eqno(2.14)$$ because of the 
terms containing Pauli matrices and therefore $H_{{\rm{e.m.}}}$ is not 
invariant under a time reversal.

We may also introduce a pseudo time reversal operator ${\cal T}$, whose
representation is 
$${\cal T}\vec A(\vec r,t){\cal T}^{-1}=-\vec A(\vec r,-t),\eqno(2.15)$$
$${\cal T}\Psi_j(t){\cal T}^{-1}=\kappa^\prime\Psi_j(-t)\eqno(2.16)$$ and
$${\cal T}\Psi_j^\dagger(t){\cal T}^{-1}=\kappa^{\prime *}
\Psi_j^\dagger(-t).\eqno(2.17)$$ with $\kappa^\prime$ another phase 
factor. It is interesting to notice that except the last two terms of (2.9), 
the rest of $H_{{\rm{e.m.}}}$ is invariant under ${\cal T}$. 
This property is very useful for the one-loop calculations in the 
next section. The Hamiltonian (2.1) is also invariant under the space 
inversion $$P\vec A(\vec r,t)P^\dagger=-\vec A(-\vec r,t)\eqno(2.18)$$ and 
$$P\Psi_j(t)P^\dagger=\eta\Psi_{j^*}(t)\eqno(2.19)$$ with $j^*$ the inversion 
image of $j$ and $\eta$ a phase factor.

The complicated dependence of $U_{ij}$ makes the perturbative 
expansion in terms of the fine structure constant rather cumbersome. 
we shall make a parabolic approximation for the rest of the paper. 
We found in I that, at the absence of the vector potential, (2.10) reduces 
to $\sum_{\vec p}\Psi_{\vec p}^\dagger E_{\vec p}\Psi_{\vec p}$ 
with $4\times 1$ column matrix $\Psi_{\vec p}$ and $E_{\vec p}$ the 
$4\times 4$ hermitian matrix.
$$E_{\vec p}=-t^\prime u_{\vec p}-tv_{\vec p}\rho_1-{1\over 2}
\lambda S(\rho_+\hat\zeta_A\cdot\vec \tau
+\rho_-\hat\zeta_B\cdot\vec \tau), \eqno(2.20)$$  
where $$\tau_1=\left(\matrix{\sigma_1&0\cr 0&\sigma_1}\right),
\tau_2=\left(\matrix{\sigma_2&0\cr 0&\sigma_2}\right),
\tau_3=\left(\matrix{\sigma_3&0\cr 0&\sigma_3}\right)\eqno(2.21)$$
and $$\rho_1=\left(\matrix{0&I\cr I&0}\right),
\rho_2=\left(\matrix{0&-iI\cr iI&0}\right), 
\rho_3=\left(\matrix{I&0\cr 0&-I}\right)\eqno(2.22)$$ 
with $$[\rho_a,\tau_b]=0.\eqno(2.23)$$ 

The kinetic energy, $\sum_{\vec p}\Psi_{\vec p}^\dagger E_{\vec p}
\Psi_{\vec p}$, is invariant under the ${\cal C}$-transformation, 
${\cal C}\Psi_{\vec p}{\cal C}^{-1}=-\rho_1\tau_3\Psi_{\vec p}$
and ${\cal C}^2=1$. This symmetry can be diagonalized through the unitary 
transformation $\Psi_{\vec p}=V\psi_{\vec p}$ with $$V={1\over\sqrt{2}}
(\rho_-+\rho_+\tau_3)(\rho_1+\rho_3)\Big(\cos{\Theta\over 4}
-i\tau_2\sin{\Theta\over 4}\Big).\eqno(2.24)$$
      
We find that $$V^\dagger E_{\vec p}V=-t^\prime u_{\vec p}
-tv_{\vec p}\rho_3\Big(\tau_3\cos{\Theta\over 2}-\tau_1\sin{\Theta\over 2}
\Big)-{1\over 2}\lambda S\tau_3,\eqno(2.25)$$ which is easily diagonalized 
and produces four energy bands:
$$\epsilon_{\vec p}=-t^\prime(u_{\vec p}+\mu)\pm\Delta_{\vec p}^{\pm}
\eqno(2.26)$$ with $$\Delta_{\vec p}^\pm=\sqrt{{1\over 4}\lambda^2S^2\pm\lambda 
Stv_{\vec p}\cos{\Theta\over 2}+t^2v_{\vec p}^2},\eqno(2.27)$$
where $\pm$ of $\Delta_{\vec p}$ labels the eigenvalue of the ${\cal C}$ 
operator. The lowest band corresponds to 
$$\epsilon_{\vec p}=-t^\prime(u_{\vec p}+\mu)-\Delta_{\vec p}^+\eqno(2.28)$$
which overlaps partially at the top with the next lowest band, i.e., 
$\epsilon_{\vec p}=-t^\prime(u_{\vec p}+\mu)-\Delta_{\vec p}^-$. In what 
follows we shall assume that only the band (2.28) is partially filled, a 
likely situation with doping conditions of $(Sr, Ba)_2YRu_{1-x}Cu_xO_6$. 
With the parabolic approximation we have 
$$u_{\vec p}=2-{l^2\over 4}(p_x^2+p_y^2)\eqno(2.29)$$ and 
$$v_{\vec p}=4-{l^2\over 4}(p_x^2+p_y^2+2p_z^2).\eqno(2.30)$$ Regarding 
$\psi_{\vec p}$ as the Fourier component of a $4\times 1$ fermion 
field $\psi(\vec r)$ in continuum, we have ${2\over {\cal N}}\sum_{\vec p}
=\int{d^3\vec p\over (2\pi)^3}$ and the "coordinate" representation 
of kinetic energy operator, (2.25), reads
$$K=\int d^3\vec r\Big[{1\over 2}\vec\nabla_\perp\psi^\dagger\cdot
\Big({1\over m^\prime}+{1\over m}\rho_3\tau_3^\prime\Big)\vec\nabla_\perp
\psi+{1\over m}{\partial\over\partial z}\psi^\dagger\rho_3
\tau_3^\prime{\partial\over\partial z}\psi$$ 
$$-\nu\psi^\dagger\rho_3\tau_3^\prime\psi-\mu_c\psi^\dagger\psi
-{1\over 2}\lambda S\psi^\dagger\tau_3\psi\Big],\eqno(2.31)$$ where 
the parameters $m$, $m^\prime$ and $\nu$ are related 
to the discrete Hamiltonian through $m^{\prime-1}=t^\prime l^2/2$, 
$m^{-1}=tl^2/2$, $\nu=4t$, $\mu_c=\mu+2t^\prime$. Replace the ordinary 
derivatives by the covariant ones, we obtain the continuum approximation 
of (2.10), i.e.
$$H_{{\rm{e.m.}}}=\int d^3\vec r{1\over 2}(\vec E_{{\rm{tr.}}}^2
+\vec B^2)+\int d^3\vec r\Big[{1\over 2}(\vec\nabla_\perp
+ie\vec A_\perp)\psi^\dagger\cdot\Big({1\over m^\prime}+
{1\over m}\rho_3\tau_3^\prime\Big)(\vec\nabla_\perp-ie\vec A_\perp)\psi$$
$$+{1\over m}({\partial\over\partial z}+ieA_z)\psi^\dagger\rho_3
\tau_3^\prime({\partial\over\partial z}-ieA_z)\psi$$ 
$$-\nu\psi^\dagger\rho_3\tau_3^\prime\psi-\mu_c\psi^\dagger\psi
-{1\over 2}\lambda S\psi^\dagger\tau_3\psi-{g_e\over 2}
\mu_B\psi^\dagger\vec\alpha\cdot\vec B\psi\Big],\eqno(2.32)$$ where 
$$\vec\alpha=V^\dagger\vec\tau V=-\rho_1\tau_1^\prime\hat\xi
-\rho_1\tau_2\hat\eta+\tau_3^\prime\hat\zeta,\eqno(2.33)$$
$$\tau_3^\prime=\tau_3\cos{\Theta\over 2}-\tau_1\sin{\Theta\over 2}
,\eqno(2.34)$$ $$\tau_1^\prime=\tau_3\sin{\Theta\over 2}
+\tau_1\cos{\Theta\over 2},\eqno(2.35)$$ $e$ is positive for holes and 
negative for electrons and we have substituted $g_e\nu_B$ for $\mu_h$.

It is well known that, the Hamiltonian (2.10) and (2.32) produce the same 
physics as long as the length scales involved are much larger than 
the lattice spacing, $l$. The Hamiltonian 
(2.32) is not bounded from below but this does no harm for the perturbative 
calculations provide the condition of the parabolic approximation is 
met. Like the Hamiltonian (2.10) the Hamiltonian (2.32) is not invariant 
under a time reversal. But the pseudo time reversal transformation leaves 
all terms except the last one (Pauli term) invariant.

The electric current operator of the system consists of three terms
$$\vec J(\vec r,\vec t)=\vec J_d(\vec r,\vec t)+\vec J_p(\vec r,\vec t)
+\vec J_s(\vec r,\vec t)\eqno(2.36)$$ where $\vec J_d$, $\vec J_p$ and 
$\vec J_s$ denote the diamagnetic, paramagnetic and spin part and their 
expressions are listed below
$$\vec J_d(\vec r,t)=-{e^2\over 2}\psi^\dagger\Big({1\over m^\prime}
+{1\over m}\rho_3\tau_3^\prime\Big)\psi\vec A_{\perp}-
{e^2\over m}\psi^\dagger\rho_3\tau_3^\prime\psi\vec A_z\hat z;\eqno(2.37)$$
$$\vec J_p(\vec r,t)=-{ie\over 2}\Big[\psi^\dagger
\Big({1\over m^\prime}+{1\over m}\rho_3\tau_3^\prime\Big)
\vec\nabla_{\perp}\psi-{\rm{h.c.}}\Big]-{ie\over m}\Big(\psi^\dagger\rho_3
\tau_3^\prime{\partial\over \partial z}\psi-{\rm{h.c.}}\Big)\hat z\eqno(2.38)
$$ and $$\vec J_s(\vec r,t)={g_e\over 2}\mu_B\vec\nabla\times
\psi^\dagger\vec\alpha\psi.\eqno(2.39)$$ 

\section{3. The Response Functions}

When the system is illuminated by a beam of light with a wave vector 
$\vec k$. The optical response is entirely determined by the transverse 
components of the retarded photon self energy tensor in the light frame, 
i.e. $$\sigma_{\alpha\beta}(\omega)=(\hat e_\alpha)_i\Pi_{ij}^R
(\omega,\vec k)(\hat e_\alpha)_j\eqno(3.1)$$ where, at zero temperature, 
$${\rm{Re}}\Pi_{ij}^R(\omega,\vec k)={\rm{Re}}\Pi_{ij}(\omega,\vec k)
\eqno(3.2)$$ and $${\rm{Im}}\Pi_{ij}^R(\omega,\vec k)={\rm{sign}}
(\omega){\rm{Im}}\Pi_{ij}(\omega,\vec k)\eqno(3.3)$$ with 
$\Pi(\omega,\vec k)$ time-ordered self energy tensor which can be 
calculated diagrammatically.
For a system invariant under the space inversion, we expect
$$\Pi_{ij}(\omega,\vec k)=\Pi_{ij}(\omega,-\vec k).\eqno(3.4)$$ 
For a system invariant under time reversal, we have
$$\Pi_{ij}(\omega,\vec k)=\Pi_{ji}(\omega,-\vec k).\eqno(3.5)$$ 
Come to the system we are considering, the time reversal invariance 
is spontaneously broken by the magnetic ordering but inversion 
symmetry is intact. The self energy tensor will acquire an antisymmetric 
part, i.e. $$\Pi_{ij}(\omega,\vec k)-\Pi_{ji}(\omega,\vec k)\neq 0,
\eqno(3.6)$$ which we shall focus our attention on for the rest of the 
article. 

The one-loop diagrams for $\Pi_{ij}(\omega,\vec k)$ are displayed 
in Fig. 2. 
\topinsert
\hbox to\hsize{\hss
	\epsfxsize=4.0truein\epsffile{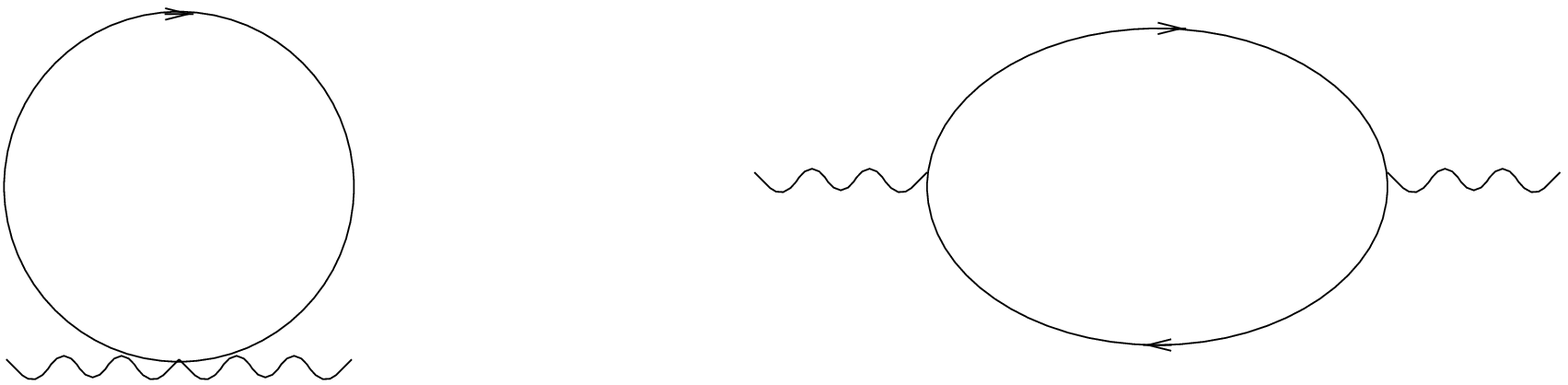}
	\hss}
\begincaption{Figure 2}
The Feynman diagrams for the photon self energy tensor.
\endcaption
\endinsert
The evaluation of them is straightforward and we tabulate 
below the results under the approximation
$\lambda S>>t,t^\prime$ $\omega \sim \lambda S$ and $\rho_hl^3<<1$ 
where $\rho_h$ is the number of itinerant holes per unit volume,
$$\rho_h=\int{d^3\vec p\over (2\pi)^3}\theta(\mu_c-\epsilon_{\vec p})
\eqno(3.7)$$ with $\theta(x)={1\over 2}(1+{\rm{sign}}(x))$. 
We divide $\Pi_{ij}(\omega,\vec k)$ into three terms, 
$$\Pi_{ij}(\omega,\vec k)=\Pi_{ij}^{(1)}(\omega,\vec k)
+\Pi_{ij}^{(2)}(\omega,\vec k)+\Pi_{ij}^{(3)}(\omega,\vec k)\eqno(3.8)$$
The $\Pi^{(1)}$ term comes from the orbital motion only and is given by
$$\Pi_{ij}^{(1)}(\omega,\vec k)=e^2\rho_h\Big[{1\over m^\prime}
(\delta_{ij}-\delta_{iz}\delta_{jz})+{1\over m}\cos{\Theta\over 2}
(\delta_{ij}+\delta_{iz}\delta_{jz})\Big]$$
$$-i\int d^3\vec re^{-i\vec k\cdot\vec r}\int_{-\infty}^{\infty}dt
e^{i\omega t}<|TJ_{p,i}(\vec r,t)J_{p,j}(0,0)|>$$
$$=e^2\rho_h\Big[{1\over m^\prime}
(\delta_{ij}-\delta_{iz}\delta_{jz})+{1\over m}\cos{\Theta\over 2}
(\delta_{ij}+\delta_{iz}\delta_{jz})\Big]$$
$$-{e^2\rho_h\over m^2}{\lambda S\over \lambda^2S^2-\omega^2}
\Big[<p_\perp^2>(\delta_{ij}-\delta_{iz}\delta_{jz})
+8<p_z^2>\delta_{iz}\delta_{jz}\Big]\sin^2{\Theta\over 2},\eqno(3.9)$$
where $$<F>\equiv{1\over \rho_h}\int{d^3\vec p\over (2\pi)^3}\theta(
-\epsilon_{\vec p})F(\vec p).\eqno(3.10)$$
The $\Pi^{(2)}$ term of (3.8) represent the cross term between the 
paramagnetic current and the spin current and is given by
$$\Pi_{ij}^{(2)}(\omega,\vec k)=-i\int d^3\vec re^{-i\vec k\cdot\vec r}
\int_{-\infty}^{\infty}dt e^{i\omega t}<|T[J_{p,i}(\vec r,t)J_{s,j}(0,0)
+J_{s,i}(\vec r,t)J_{p,j}(0,0)]|>$$ $$=i[\epsilon_{imn}k_m\hat\zeta_n
\Lambda_j(\omega,\vec k)-\epsilon_{jmn}k_m\hat\zeta_n\Lambda_i
(\omega,\vec k)]\eqno(3.11)$$ where
$$\Lambda_j(\omega,\vec k)=g_e{\mu_B e\rho_h\over 2m\omega}
\Big[\Big({m\over m^\prime}\cos{\Theta\over 2}+{\lambda^2S^2\cos^2
{\Theta\over 2}-\omega^2\over \lambda^2S^2-\omega^2}\Big)k_j(1-\delta_{jz})$$
$$+2{\lambda^2S^2\cos^2{\Theta\over 2}-\omega^2\over \lambda^2S^2-
\omega^2}k_z\delta_{jz}\Big].\eqno(3.12)$$ Finally, 
the $\Pi^{(3)}$ term is related to the spin susceptibility tensor 
$\chi_{ij}(\omega)$ via $$\Pi_{ij}^{(3)}(\omega,\vec k)=
\int_{-\infty}^{\infty}dt e^{i\omega t}<|TJ_{s,i}(\vec r,t)J_{s,j}(0,0)|>
=\epsilon_{imn}\epsilon_{jm^\prime n^\prime}
k_mk_{m^\prime}\chi_{nn^\prime}(\omega),\eqno(3.13)$$ where
$$\chi_{ij}(\omega)=R_{i\mu}R_{j\nu}\chi_{\mu\nu}(\omega)
\eqno(3.14)$$ with the indices $\mu$, $\nu$ referring to axes of 
the spin frame ($\mu,\nu=\xi,\eta,\zeta$) and $R_{i\mu}$ 
the elements of the orthogonal matrix 
$$R=\left(\matrix{\hat x\cdot\hat\xi&\hat x\cdot\hat\eta&
\hat x\cdot\hat\zeta\cr \hat y\cdot\hat\xi&\hat y\cdot\hat\eta&
\hat y\cdot\hat\zeta\cr \hat z\cdot\hat\xi&\hat z\cdot\hat\eta&
\hat z\cdot\hat\zeta\cr}\right).\eqno(3.15)$$ The nonvanishing components 
of $\chi_{\alpha\beta}(\omega)$ are
$$\chi_{\xi\xi}(\omega)=-{1\over 2}g_e^2 \rho_h\mu_B^2
{\lambda S\cos^2{\Theta\over 2}\over\lambda^2S^2-\omega^2},\eqno(3.16)$$ 
$$\chi_{\eta\eta}(\omega)=-{1\over 2}g_e^2 \rho_h\mu_B^2
{\lambda S\over\lambda^2S^2-\omega^2},\eqno(3.17)$$ 
$$\chi_{\xi\eta}(\omega)=-\chi_{\eta\xi}(\omega)
=-{i\over 2}g_e^2 \rho_h\mu_B^2{\omega\cos{\Theta\over 2}
\over\lambda^2S^2-\omega^2},\eqno(3.18)$$ 
$$\chi_{\zeta\zeta}(\omega)=-{1\over 2}g_e^2 \rho_h\mu_B^2{\lambda S
\over \lambda^2S^2-\omega^2}\sin^2{\Theta\over 2}.\eqno(3.19)$$ 
We observe that 1) The antisymmetric part comes only from $\Pi^{(2)}$ 
and $\Pi^{(3)}$ terms (the statement is true without the approximation
$\lambda S>>t,t^\prime$ and $n_hl^3<<1$), while $\Pi^{(1)}$ remains 
symmetric, a consequence of ${\cal T}$-invariance. 2). With the large double 
exchange approximation, the symmetric part of $\Pi_{ij}(\omega,\vec k)$
is dominated by the first term of (3.9) and the antisymmetric terms are 
smaller than this term by a factor of the order of $k^2/m\lambda S$.
For $\omega$ not at the pole, $\pm\lambda S$, we have 
$\Pi_{ij}^R(\omega,\vec k)=\Pi_{ij}(\omega,\vec k)$ and from this 
the expression of $\sigma_{\alpha\beta}(\omega,\vec k)$ follows.

The propagation mode of a photon is determined by 
$$[(\omega^2-k^2)\delta_{\alpha\beta}-\sigma_{\alpha\beta}
(\omega,\vec k)]\phi_\beta=0\eqno(3.20)$$ with $\vec\phi$ the polarization 
vector, $\vec\phi\perp\vec k$. The $T$-violation implies 
that the eigenmodes are elliptically polarized instead of linearly 
polarized and the system possesses the optical activity.

Consider a special circumstance, in which $\vec k$, $\hat\zeta$ 
and $\hat z$ are all parallel, and the magnetic order is purely 
ferromagnetic. The self energy tensor $\sigma_{\alpha\beta}$
reduces to 
$$\sigma_{\alpha\beta}(\omega,\vec k)=e^2\rho_h\Big({1\over m^\prime}
+{1\over m}\Big)\delta_{\alpha\beta}-{i\over 2}k^2\epsilon_{\alpha\beta}
g_e^2\rho_h\mu_B^2{\omega\over \lambda^2S^2-\omega^2},\eqno(3.21)$$
where we have shown explicitly the symmetric and antisymmetric parts 
of $\sigma_{\alpha\beta}(\omega,\vec k)$ and only the leading order 
terms are retained in each part. The eigenmode equation (3.20) are 
reduced to 
$$\Big[(\omega^2-\omega_p^2)\delta_{\alpha\beta}-n^2\omega^2
\Big(1-{i\over 2}\epsilon_{\alpha\beta}g_e^2\rho_h\mu_B^2{\omega
\over\lambda^2S^2-\omega^2}\Big)\Big]\phi_\beta=0,\eqno(3.22)$$
where we have introduced the index of refraction, $n(\omega)$ 
via $k=n(\omega)\omega$ and the plasma frequency $\omega_p^2
=e^2\rho_h({1\over m^\prime}+{1\over m}\cos{\Theta\over 2})$. 
Clearly, the eigenmodes of (3.20) are two circularly polarized photon with 
opposite senses. The corresponding indices of refraction differ by an 
amount $$\Delta n(\omega)={1\over 2}\sqrt{\epsilon_0}g\mu_B^2
{\omega \over \lambda^2S^2-\omega^2}\eqno(3.23)$$ with $\epsilon_0=
1-{\omega_p^2\over \omega^2}$, which gives rise to a Farady angle
$$\Delta\theta=\pi\Delta n(\omega)\eqno(3.24)$$ per wavelength. 
For $Sr_2YRu_{1-x}Cu_xO_6$, we have $\rho_h=2x/v$ with $v=l^3/4=
136\AA^3$. Assuming $g_e=2$, $\Theta=0$ and ${1\over m}+{1\over {m^\prime}}
={1\over m_e}$,  we obtain:
$$\Delta\theta=0.62\times10^{-4}{\sqrt{\epsilon_0}x\omega\over
\lambda^2 S^2-\omega^2}\eqno(3.25)$$ with $\lambda S$, $\omega$ in ev and 
$\Delta\theta$ in radians.  

There are two additional contribution to the $T$-violation signal which 
are not included in the Hamiltonian (2.32) and their magnitudes turn 
out to be comparable to (3.25). 

{\it{1. The secondary magnetooptical effect}}
The aligned ion-spins will produced a net magnetic field, which gives rise 
to the ordinary magnetooptical effect [8]. For a magnetic field $\vec B=
B\hat z$ and $\vec k=k\hat z$, the Farady angle reads 
$$\Delta\theta^\prime=\pi{\omega_p^2\omega_B\over\sqrt{\epsilon_0}\omega^3}
\eqno(3.26)$$ with $\omega_B=eB({1\over m^\prime}+{1\over m}\cos
{\Theta\over 2})$ the cyclotron frequency (See appendix A for the derivation 
with a general anisotropy). For a spheroidal sample with $\vec M$ parallel 
to its symmetry axis, the average
magnetic field is $$\vec B=\gamma\vec M$$ with $\vec M$ the 
magnetization given by $$\vec M={g_i\mu_BS\over v}\cos{\Theta\over 2}$$ 
with $g_i$ the Lande factor of the core ion, where the contribution from 
the itinerant holes has been neglected on account of low dopping condition. 
The factor $\gamma$ changes from zero in the 
limit of a thin disc perpendicular to the field to one in the 
limit of a long rod along the field. Assuming $g_i=2$, $S={3\over 2}$,
$\Theta=0$ and ${1\over m^\prime}+{1\over m}\sim{1\over m_e}$, we find 
the corresponding Farady angle $$\Delta\theta^\prime
=1.89\times10^{-3}{\gamma x\over\sqrt{\epsilon_0}\omega^3}.\eqno(3.27)$$
Though this effect could be considerably larger that (3.25) but the double 
exchange effect may still be detectable because of its resonance 
nature. 

{\it{2. Spin-orbit coupling}}
It is interesting to know there is a relativistic term which 
turns out to to comparable with (3.25). The physical reason is the same 
as that for the spin-orbital coupling of atomic levels. In what follows, 
we give a classical argument [9] and defer the quantum mechanical treatment 
to the appendix B. For an electron(hole) moving in an external 
electromagnetic field $(\vec E, \vec B)$ at velocity $\vec v(v<<1)$, 
The spin precession equation reads
$${d\vec s\over dt}=\vec s\times({e\over m_e}\vec B^\prime-\vec\omega_T),
\eqno(3.28)$$ where $\vec B^\prime$ is the magnetic field in the rest 
frame of the electron(hole) and $\omega_T$ is the angular velocity of 
Thomas precession. We have set $g_e=2$ in (3.28) since we do not know
how to renormalize the bare value in a medium for this effect. The equation 
(3.28) can be derived with a Hamiltonian term 
$$H^\prime=-(2e\mu_B\vec B^\prime-\vec\omega_T)\cdot\vec s.\eqno(3.29)$$
To the first power of the velocity, we have
$$\vec B^\prime=\vec B-\vec v\times\vec E\eqno(3.30)$$ 
and $$\vec\omega_T=-{1\over 2}\vec v\times\vec a=-\mu_B\vec v\times\vec E
\eqno(3.31)$$ with $\vec a={e\over m_e}\vec E$ to the order of the 
approximation. Substituting (3.30) and (3.31) into (3.29) and express 
$\vec v$ in terms of the canonical momentum and the gauge potential,
$\vec v={1\over m_e}(\vec p-e\vec A)$, we obtain
$$H^\prime=-2\mu_B\vec s\cdot\vec B-{e\over 2m_e}\vec s\cdot\vec E
\times(\vec p-e\vec A)\eqno(3.32)$$ as an improved Pauli term.
The term $\vec E\times\vec A$ pick up the information of the violation of 
the time reversal invariance with the corresponding Farady 
angle per wavelength given by
$$\Delta\theta^{\prime\prime}={\pi e^2\rho_h^2\over 2\sqrt{\epsilon_0}
m_e^2\omega}\cos{\Theta\over 2}=0.62\times 10^{-4}{x\over 
\sqrt{\epsilon_0}\omega}\cos{\Theta\over 2}.\eqno(3.33)$$

\section{4. Discussions}

In the previous sections, we analyzed the photon self energy tensor in 
a canting magnetic ordering, focusing the signal of the $T$-violation.
There are three contributions to the optical activity, eqs. (3.25), (3.27)
and (3.28). Although the equivalent magnetic field due to the double 
exchange is huge, the magnitude of the Farady angle turns out rather small. 
The reason for this is that the double exchange acts only on the carrier
spins. The orbital motion, neglecting the weak magnetic field produced 
by the ordered ion spins, does not give rise any $T$-violation signal, 
on account of the ${\cal T}$ invariance defined in eqs.(2.15-17). Such an 
small effect lies at the border of the detection precision for a sample of 
one wavelength thick, as inferred from the delicate circular dichroism 
experiment, designed to test the anyon model of the high $T_C$ 
superconductivity [10]. On the other hand, the Farady angle associated 
with $T$-violation accumulates as the photon traverses back and forth 
inside the sample [9]. The effect may be enhanced through multiple 
reflections at the inner surface of the sample.

For the double exchange energy $\lambda S$ comparable with the photon 
energy, $\omega$, resonance character will show up in the contribution 
from the spin susceptibility, (3.25). A closer look will reveal an 
interesting point which has been watered down in the leading order 
of the strong double exchange approximation made in the eqs. (3.9-19). 
According to the Hamiltonian (2.32), the transition involved in the 
photon self energy includes both ${\cal C}$ preserving ones and 
${\cal C}$ flipping ones. In the former case, the energy denominator
for the transition reads
$$2\Delta_{\vec p}^+\pm\omega=\lambda S\pm\omega+2tv_{\vec p}
\cos{\Theta\over 2}+O\Big({t^2\over\lambda S}\Big),\eqno(4.1)$$
while in the latter case, the corresponding energy denominator 
is $$\Delta_{\vec p}^++\Delta_{\vec p}^-\pm\omega=\lambda S\pm
\omega+{t^2\over \lambda S}v_{\vec p}\cos{\Theta\over 2}.\eqno(4.2)$$
Upon integration over $\vec p$ within the Fermi sea, the resonance 
signal will be slightly smeared. such a smearing effect for the 
${\cal C}$ flipping transition is of one order higher than that 
for the ${\cal C}$ preserving transition. The transition contributing 
to $\chi_{\xi\eta}(\omega)$ is entirely ${\cal C}$ flipping type and 
the resonance signal in the Farady angle (3.25) should be fairly sharp. 
This feature is also independent of the parabolic approximation we made 
in the calculation.

The frequency dependences of the other two contributions are rather 
robust. The former involves the transition among different Landau 
levels caused by the real local magnetic field of the ordered magnetic 
moments of the ion cores. This field is in general fairly weak, of 
the order of 1000 Gauss, the level spacing associated with this 
magnetic field is tiny in comparison to the photon energy, $\omega$ 
and the energy denominator is dominated by the photon energy. The 
relativistic term (3.28), however, is the consequence of the virtual 
transition between the Dirac sea and the Fermi sea and the corresponding 
energy denominator is dominated by twice of the rest energy of an electron.

Experimentally, superconductivity was also discovered the double perovskite 
samples. A natural question arise as to the effect of the long range 
order on the optical activity calculated with normal electrons. For the 
photon in the visible region, its energy are much higher than the gap-energy 
and correction to and $\Pi_{ij}^{(2)}(\omega,\vec k)$, 
$\Pi_{ij}^{(3)}(\omega,\vec k)$ is of higher orders. Only thing we need 
to watch out is whether there is small term coming from 
$\Pi_{ij}^{(1)}(\omega,\vec k)$. The long range order can be introduced 
by adding a pairing term to the Haniltonian (2.32). The total Hamiltonian 
reads now, $${\cal H}=H+\sum_{\vec q}\beta_{\vec q}^\dagger\beta_{\vec q},
\eqno(4.3)$$ where $a_{\vec p}$, $a_{\vec p}^\dagger$ stands for the 
annihilation and creation operators of the itinerant holes, and 
$\beta_{\vec q}$, $\beta_{\vec q}^\dagger$ stands for composite 
boson operators which represent Cooper pairs, i.e., 
$$\beta_{\vec q}={1\over\sqrt{{\cal N}}}\mathop{{\sum}'}_{\vec p} 
g_{\vec p,\vec q}a_{\vec p+{\vec q\over 2}}a_{-\vec p+{\vec q\over 2}}
\eqno(4.4)$$ with $\mathop{{\sum}'}_{\vec p}$ extends over half of the 
Brillouin zone only. The pairing wave function, $g_{\vec p,\vec q}$, 
is odd under the space inversion, i.e.
$$g_{\vec p,\vec q}=-g_{-\vec p,\vec q}\eqno(4.5)$$ on account of the 
anticommutation relation among $a_{\vec p}$'s. The superconductivity 
is implemented through the condensation of the pairing operator of 
zero total momentum, i.e.
$$<S|\beta_{\vec q=0}|S>=\sqrt{{\cal N}}B,\eqno(4.6)$$ where $B$ is the 
long range order parameter. The wave function of the ground state is then
$$|S>=\mathop{{\prod}'}_{\vec p}(\cos\theta_{\vec p}-e^{-i{\gamma_{\vec p}}}
\sin\theta_{\vec p}a_{\vec p}^\dagger a_{-\vec p}^\dagger)|0>,\eqno(4.7)$$
where $\cos2\theta_{\vec p}={\epsilon_{\vec p}\over {\cal E}_{\vec p}}$ 
and $\sin2\theta_{\vec p}={|\delta_{\vec p}|\over {\cal E}_{\vec p}}$
with ${\cal E}_{\vec p}=\sqrt{\epsilon_{\vec p}^2+|\delta_{\vec p}|^2}$ 
and $|\delta_{\vec p}|e^{i\gamma_{\vec p}}=Bg_{\vec p,\vec q}\perp_
{\vec q=0}$. Evidently, both the standard time reversal $T$ and the pseudo
time reversal ${\cal T}$ do not leave the ground state (4.7) unchanged. 
On the other hand, The following antiunitary transformation
$${\cal T}^\prime a_{\vec p}{\cal T}^{\prime -1}=ie^{i\gamma_{\vec p}}
a_{\vec p}\eqno(4.8)$$ leaves both the pairing Hamiltonian as well as 
the ground state (4.7) invariant. This, together with the invariance 
under the modified space inversion in I, ${\cal P}=Pe^{{\pi\over 2}Q}$, 
implies that $\Pi_{ij}^{(1)}(\omega,0)=\Pi_{ji}^{(1)}(\omega,0)$ and the 
T-violation signal is suppressed relative to $\Pi_{ij}^{(1)}(\omega,0)$ at 
least by a factor $k^2/p_F^2$ with $p_F$ the Fermi-momentum. The very fact 
that they must vanish as the gap energy vanishes pushes it much smaller 
than the magnitude of $\chi_{\xi\eta}(\omega)$. Therefore the estimate 
of the T-violation signals (3.25) and (3.28) with normal electrons applies 
to the superconducting phase as well.

As was pointed out in [11] in the context of Raman scattering, 
the way we introduce the gauge invariant coupling, eqs. (2.11) and (3.32), 
requires that the photon frequency in the energy denominators for the 
transition to other bands not covered by the model Hamiltonian (1.1) can be 
dropped. This may be rather marginal for real system. Also the impurity 
effect, which may be relevant to the real system has been neglected. 

\section{Acknowledgments}

Hai-cang Ren's work is supported in part by U. S. Department of Energy 
under Grant DE-FG02-91ER40651, Task B and M. K. Wu's work is supported 
in part by the ROC National Science Council grant \#NSC87-0511-M-007-004.

\section{Appendix A}

For the ordinary magneto-optic effect, we consider only the conduction 
band. The effective Hamiltonian with the parabolic approximation reads
$$H={1\over 2}\int d^3\vec r \Big[{1\over 2m_1}({\partial\over\partial x}
+ieA_x)\psi^\dagger({\partial\over\partial x}-ieA_x)\psi$$
$$+{1\over 2m_2}({\partial\over\partial y}
+ieA_y)\psi^\dagger({\partial\over\partial y}-ieA_y)\psi
+{1\over 2m_3}({\partial\over\partial z}
+ieA_z)\psi^\dagger({\partial\over\partial z}-ieA_z)
\psi \Big],\eqno(A.1)$$ where we have introduced general anisotropy and the 
symmetry of (A.1) is orthohmobic. In our case, which is tetragonal
$${1\over m_1}={1\over m_2}={1\over m^\prime}+{1\over m}\cos{\Theta\over 2}
\eqno(A.2)$$ and $${1\over m_3}={2\over m}\cos{\Theta\over 2}.\eqno(A.3)$$
The gauge potential $\vec A(\vec r,t)$ is chosen to produce a static 
and homogeneous magnetic field, 
$$\vec B=\vec\nabla\times\vec A(\vec r,t)\eqno(A.2)$$ and a time dependent 
and homogeneous electric field,
$$\vec E=-{\partial\vec A\over \partial t}.\eqno(A.3)$$

The electric current operator reads $$J_a=-{ie\over m_a}\psi^\dagger
({\partial\over\partial x_a}-ieA_a)\psi +{{\rm{h.c.}}}\eqno(A.4)$$ 
with $a=1,2,3$, and the charge density operator is
$$J_0=e\psi^\dagger\psi.\eqno(A.5)$$ 

Let $|>$ denote the ground state of (A.1). It follows from the translation 
invariance of $\vec E$ and $\vec B$ and the gauge invariance of (A.1) that 
the expectation values $<|\vec J(\vec r,t)|>$ and $<|J_0(\vec r,t)|>$ are 
coordinate independent. Introducing the operators
$$\vec{\cal J}={1\over \Omega}\int d^3\vec r\vec J(\vec r,t)\eqno(A.6)$$ 
and $${\cal J}_0={1\over \Omega}\int d^3\vec rJ_0(\vec r,t),\eqno(A.7)$$ 
we have $$\vec j\equiv <|\vec J(\vec r,t)|>=<|\vec{\cal J}|>\eqno(A.8)$$ 
and $$e\rho_h\equiv <|J_0(\vec r,t)|>=<|{\cal J}_0|>\eqno(A.9)$$
with $\rho_h$ the number density of the carriers (holes in our case).
In Heisenberg representation, the operator equation of motion for 
$\vec{\cal J}$ is identical in form to the classical one, i.e.
$${d{\cal J}_a\over dt}={e^2\over m_a}{\cal J}_0E_a+{e^2\over m_a}
\sum_{b,c}\epsilon_{abc}{\cal J}_bB_c\eqno(A.10)$$ and 
${d\over dt}{\cal J}_0=0$ on account of the charge conservation.
Taking the expectation value of (A.10) with respect to $|>$, we find that
$${dj_a\over dt}={e^2\rho_h\over m_a}E_a+{e\over m_a}\sum_{b,c}
\epsilon_{abc}j_bB_c.\eqno(A.11)$$ The anisotropy can be scaled away as:
$$j_a={1\over\sqrt{m_a}}\hat j_a,\eqno(A.12)$$ $$E_a=\sqrt{m_a}\hat E_a
\eqno(A.13)$$ and $$B_c=\sqrt{m_am_b}\hat B_c\eqno(A.14)$$ with 
$a,b,c$ a permutation of $1,2,3$. Then 
$${d\hat j_a\over dt}=e^2\rho_h\hat E_a+e\sum_{b,c}\epsilon_{abc}
\hat j_b\hat B_c.\eqno(A.15)$$ For $\hat E_a=\hat E_a(\omega)e^{-i\omega t}$, 
we find $\hat j_a=\hat j_a(\omega)e^{-i\omega t}$ with 
$$\hat j_a(\omega)=\sum_b\hat\sigma_{ab}(\omega)\hat E_b(\omega),\eqno(A.16)$$
where $$\hat\sigma_{ab}(\omega)=i{e^2\rho_h\omega\over \omega^2-\omega_B^2}
\Big(\delta_{ab}-{e^2\over \omega^2}\hat B_a\hat B_b\Big)-{e^3\rho_h\over
\omega^2-\omega_B^2}\sum_c\epsilon_{abc}\hat B_c\eqno(A.17)$$ with
$\omega_B=e\sqrt{{B_1^2\over m_2m_3}+{B_2^2\over m_3m_1}+{B_3^2\over m_1m_2}}$
the cyclotron frequency. 
Undoing the scale transformations (A.12-14), we obtain the complex 
conductivity tensor, $\sigma_{ab}(\omega)={1\over\sqrt{m_am_b}}
\hat\sigma_{ab}(\omega)$. Substituting into the standard formula, we end up 
with the dielectric tensor with respect to the lattice frame
$$\varepsilon_{ab}(\omega)=\delta_{ab}-{e^2\rho_h\over \sqrt{m_am_b}
(\omega^2-\omega_B^2)}\Big(\delta_{ab}-{e^2\over\omega^2}\hat B_a\hat B_b
+i{e\over\sqrt{m_am_b}\omega}\sum_c\epsilon_{abc}B_c\Big).\eqno(A.18)$$

\section{Appendix B}

In this appendix, we shall derive the improved Pauli term (3.32) by means 
of Foldy-Wouthuysen transformation of Dirac equation. The single 
electron Dirac equation in an ionic crystal and in an external 
radiation field reads
$$i{\partial\psi\over \partial t}=H\psi,\eqno(B.1)$$ with
$$H=-i\rho_1\vec\tau\cdot(\vec\nabla-ie\vec A)+eA_0+\vec V(\vec r)\cdot
\tau+\rho_3m_e,\eqno(B.2)$$ where $(A_0,\vec A)$ is the four component 
gauge potential with $A_0$ containing the self-consistent electrostatic 
potential of ion cores and electron clouds, $V(\vec r)$ is the exchange 
part of the self-consistent potential, $\psi(\vec r)$ is a $4\times 1$ 
spinor wave function and $\rho_i$ and $\tau_i$ are the same $4\times 4$ 
matrices introduced in the section 2 (in terms of the standard notation, 
$\rho_3=\beta$, $\rho_1\vec\tau=\vec\alpha$). Equation (B.1) can be regarded 
a relativistic Hartree-Fock equation with a filled Dirac sea of negative 
energy levels and a Fermi sea of the positive levels. As we shall see, 
to the order we are interested, the details of $A_0(\vec r)$ and 
$V(\vec r)$ are unimportant provide that their magnitudes are 
nonrelativistic and they produce the same conducting band given by (2.28).

For a nonrelativistic electron, the lower $2\times 1$ spinor of $\psi$ 
is suppressed by a factor $\sim v/c$ relative to the upper $2\times 1$ 
spinor. Following Foldy and Wouthuysen [12], we decompose the Hamiltonian 
(B.2) into three part, $$H=\rho_3m_e+{\cal E}+{\cal O},\eqno(B.3)$$ where 
$${\cal E}=eA_0(\vec r)+\rho_3\vec V(\vec r)\cdot\tau\eqno(B.4)$$
$${\cal O}=-i\rho_1\vec\tau\cdot(\vec\nabla-ie\vec A(\vec r)).\eqno(B.5)$$
Among sixteen basic $4\times 4$ matrices, 1, $\rho_i$, $\tau_i$, 
$\rho_i\tau_j$ ($i=1,2,3,j=1,2,3$), $\rho_1$, $\rho_2$, $\rho_1\tau_i$ 
and $\rho_2\tau_i$ couple to the upper $2\times 1$ spinor and the lower 
one and will be referred to as odd operator, the rest of them does not 
couple the upper $2\times 1$ spinor and the lower one and will be referred 
to as odd operator. Therefore $\rho_3m_e$ and ${\cal E}$ is even 
while ${\cal O}$ is odd. A systematic nonrelativistic approximation of 
a Dirac equation amounts to a set of successive unitary transformations 
$$\psi^\prime=e^{-iS}\psi\eqno(B.6)$$ and 
$$H^\prime=H-ie^{iS}{\partial\over \partial t}e^{-iS},\eqno(B.7)$$
which eliminates the odd operators to the desired order of 
${1\over m_e}$. For the Dirac Hamiltonian (B.2) and to the order 
${1\over m_e}$, the following three transformations serves the purpose:
$$e^{-iS}=e^{-iS_3}e^{-iS_2}e^{-iS_1}\eqno(B.8)$$ with 
$$S_1=-i{1\over 2m_e}\rho_3{\cal O}\eqno(B.9)$$ 
$$S_2={1\over 4m_e^2}(\dot{\cal O}-i[{\cal O},{\cal E}])\eqno(B.10)$$ and 
$$S_3={i\over 8m_e^3}\rho_3\Big({4\over 3}{\cal O}^3+\ddot{\cal O}
+i[{\cal E},\dot{\cal O}-i[{\cal O},{\cal E}]]-i[\dot{\cal O},{\cal E}]
-i[{\cal O},\dot{\cal E}]\Big).\eqno(B.11)$$ The transformed Hamiltonian 
reads $$H^\prime=\rho_3m_e+eA_0-{1\over 2m_e}\rho_3(\vec\nabla-ie\vec A)^2
-{e\over m_e}\rho_3\vec\tau\cdot\vec B$$ $$-{ie\over 4m_e^2}\vec\tau\cdot
\vec\nabla\times\vec E+{ie\over 4m_e^2}\vec\tau\cdot\vec E\times
(\vec\nabla-ie\vec A)-{e\over 8m_e^2}\vec\nabla\cdot\vec E$$
$$+{e\over 8m_e^2}\rho_3\{\vec\tau\cdot(\vec\nabla-ie\vec A), 
\{\vec\tau\cdot(\vec\nabla-ie\vec A),\vec V(\vec r)\cdot\vec\tau\}\}
,\eqno(B.12)$$ where $\vec E=-{\partial\vec A\over\partial r}
-\vec\nabla A_)$ and $\vec B=\vec\nabla\times\vec A$. If we neglect the 
local magnetic field produced by the spin ordering, $\vec A$ is entirely 
due to the radiation field, so is $\vec B$. The electric field $\vec E$,
however, can be decomposed into a static field caused by the ionic potential
and the radiation field. The Hamiltonian is a quadratic function of the 
radiation field, $\vec A$, $\vec E_{{\rm{rad.}}}$ and $\vec B$. For a 
nonrelativistic Fermi sea of plasma frequency $\omega_p$ and a photon of 
frequency $\omega \sim \omega_p$, $k\sim \omega$. With 
$|\vec V|\sim \omega_p$ and $eA_0(\vec r)\sim \omega_p$, it is easy 
to see that the leading contributions to the antisymmetric part of the 
self energy function come from the second order perturbation of the 
Pauli term and the first order perturbation of the term
$$-{e^2\over 4m_e^2}\vec\tau\cdot\vec A\times\vec E_{{\rm{rad.}}}
\eqno(B.13)$$ of (B.12). The latter gives rise to the Farady angle 
(3.28). 

\references
\ref{1.}{C. Zener, Phys. Rev., {\bf 82}, 403 (1951).}
\ref{2.}{P. W. Anderson and H. Hasegawa, Phys. Rev., {\bf 100}, 675 (1955).}
\ref{3.}{A. J. Mills, P. B. Littlewood and B. I. Shraiman, Phys. Rev. Lett.,
{\bf 74}, 5144 (1995) and the references therein.}
\ref{4.}{M. K. Wu, et al., Z. Phys., {\bf 102}, 37 (1997).}
\ref{5.}{P. G. de Gennes, Phys. Rev. {\bf 118}, 141 (1960).}
\ref{6.}{M. K. Wu, et al., submitted to Phys. Rev. Lett.}
\ref{7.}{H. C. Ren and M. K. Wu, submitted to Phys. Rev. B}
\ref{8.}{L. D. Landau and E. M. Lifshitz, {\it{Elictrodynamics of 
Continuous Media}}, Trans. J. B. Sykes and J. S. Bell, Pergamon Press, 
1960.}
\ref{9.}{J. D. Jackson, {\it{Classical Electrodynamics}}(Chapter 11), 
2nd ed., John Wiley \& Sons, Inc., (1975).}
\ref{10.}{K. B. Lyons, et. al., Phys. Rev. Lett., {\bf 64}, 2949 (1990);
T. W. Lawrence, et. al., {\it{ibid}}, {\bf 69}, 1439 (1992).}
\ref{11.}{A. A. Abrikosov and V. M. Genkin, Zh. Eksp. Teor. Fiz. 
{\bf 65}, 842 (1973). (Sov. Phys. JEPT, {\bf 38}, 417 (1974)).}
\ref{12.}{L. L. Foldy and S. A. Wouthuysen, Phys. Rev., {\bf 78}, 29 (1950).}
\vfill\eject
\end

\bye

%% file: gb.tex
%
%
\headline{\hfil \folio}
\hoffset=0.5truein
\hsize=5.5truein
\vsize=8truein
%
\catcode`@=11                           
\newskip\ttglue
\def\ninefonts{%
   \global\font\ninerm=cmr9%
   \global\font\ninei=cmmi9%
   \global\font\ninesy=cmsy9%
   \global\font\nineex=cmex10%
   \global\font\ninebf=cmbx9%
   \global\font\ninesl=cmsl9%
   \global\font\ninett=cmtt9%
   \global\font\nineit=cmti9%
   \skewchar\ninei='177%
   \skewchar\ninesy='60%
   \hyphenchar\ninett=-1%
   \moreninefonts
   \gdef\ninefonts{\relax}}%
\def\moreninefonts{\relax}                      


\def\elevenfonts{%
   \global\font\elevenrm=cmr10 scaled \magstephalf%
   \global\font\eleveni=cmmi10 scaled \magstephalf%
   \global\font\elevensy=cmsy10 scaled \magstephalf%
   \global\font\elevenex=cmex10%
   \global\font\elevenbf=cmbx10 scaled \magstephalf%
   \global\font\elevensl=cmsl10 scaled \magstephalf%
   \global\font\eleventt=cmtt10 scaled \magstephalf%
   \global\font\elevenit=cmti10 scaled \magstephalf%
   \global\font\elevenss=cmss10 scaled \magstephalf%
   \skewchar\eleveni='177%
   \skewchar\elevensy='60%
   \hyphenchar\eleventt=-1%
   \moreelevenfonts
   \gdef\elevenfonts{\relax}}%
\def\moreelevenfonts{\relax}

\def\twelvefonts{%
   \global\font\twelverm=cmr10 scaled \magstep1%
   \global\font\twelvei=cmmi10 scaled \magstep1%
   \global\font\twelvesy=cmsy10 scaled \magstep1%
   \global\font\twelveex=cmex10 scaled \magstep1%
   \global\font\twelvebf=cmbx10 scaled \magstep1%
   \global\font\twelvesl=cmsl10 scaled \magstep1%
   \global\font\twelvett=cmtt10 scaled \magstep1%
   \global\font\twelveit=cmti10 scaled \magstep1%
   \global\font\twelvess=cmss10 scaled \magstep1%
   \skewchar\twelvei='177%
   \skewchar\twelvesy='60%
   \hyphenchar\twelvett=-1%
   \moretwelvefonts
   \gdef\twelvefonts{\relax}}%
\def\moretwelvefonts{\relax}                    

\def\fourteenfonts{%
   \global\font\fourteenrm=cmr10 scaled \magstep2%
   \global\font\fourteeni=cmmi10 scaled \magstep2%
   \global\font\fourteensy=cmsy10 scaled \magstep2%
   \global\font\fourteenex=cmex10 scaled \magstep2%
   \global\font\fourteenbf=cmbx10 scaled \magstep2%
   \global\font\fourteensl=cmsl10 scaled \magstep2%
   \global\font\fourteenit=cmti10 scaled \magstep2%
   \global\font\fourteenss=cmss10 scaled \magstep2%
   \skewchar\fourteeni='177%
   \skewchar\fourteensy='60%
   \morefourteenfonts
   \gdef\fourteenfonts{\relax}}%
\def\morefourteenfonts{\relax}                  


\def\tenmibfonts{
   \global\font\tenmib=cmmib10%
   \global\font\tenbsy=cmbsy10%
   \skewchar\tenmib='177%
   \skewchar\tenbsy='60%
   \gdef\tenmibfonts{\relax}}

\def\elevenmibfonts{
   \global\font\elevenmib=cmmib10 scaled \magstephalf%
   \global\font\elevenbsy=cmbsy10 scaled \magstephalf%
   \skewchar\elevenmib='177%
   \skewchar\elevenbsy='60%
   \gdef\elevenmibfonts{\relax}}

\def\twelvemibfonts{
   \global\font\twelvemib=cmmib10 scaled \magstep1%
   \global\font\twelvebsy=cmbsy10 scaled \magstep1%
   \skewchar\twelvemib='177%
   \skewchar\twelvebsy='60%
   \gdef\twelvemibfonts{\relax}}

\def\fourteenmibfonts{
   \global\font\fourteenmib=cmmib10 scaled \magstep2%
   \global\font\fourteenbsy=cmbsy10 scaled \magstep2%
   \skewchar\fourteenmib='177%
   \skewchar\fourteenbsy='60%
   \gdef\fourteenmibfonts{\relax}}

\def\mib{
   \tenmibfonts%
   \textfont0=\tenbf\scriptfont0=\sevenbf%
   \scriptscriptfont0=\fivebf%
   \textfont1=\tenmib\scriptfont1=\seveni%
   \scriptscriptfont1=\fivei%
   \textfont2=\tenbsy\scriptfont2=\sevensy%
   \scriptscriptfont2=\fivesy}%

\def\ninepoint{\ninefonts               
   \def\rm{\fam0\ninerm}%
   \textfont0=\ninerm\scriptfont0=\sevenrm\scriptscriptfont0=\fiverm
   \textfont1=\ninei\scriptfont1=\seveni\scriptscriptfont1=\fivei
   \textfont2=\ninesy\scriptfont2=\sevensy\scriptscriptfont2=\fivesy
   \textfont3=\nineex\scriptfont3=\nineex\scriptscriptfont3=\nineex
   \textfont\itfam=\nineit\def\it{\fam\itfam\nineit}%
   \textfont\slfam=\ninesl\def\sl{\fam\slfam\ninesl}%
   \textfont\ttfam=\ninett\def\tt{\fam\ttfam\ninett}%
   \textfont\bffam=\ninebf
   \scriptfont\bffam=\sevenbf
   \scriptscriptfont\bffam=\fivebf\def\bf{\fam\bffam\ninebf}%
   \def\mib{\relax}%
   \tt\ttglue=.5emplus.25emminus.15em
   \normalbaselineskip=11pt
   \setbox\strutbox=\hbox{\vrule height 8pt depth 3pt width 0pt}%
   \normalbaselines\rm\singlespaced}%

\def\tenpoint{
   \def\rm{\fam0\tenrm}%
   \textfont0=\tenrm\scriptfont0=\sevenrm\scriptscriptfont0=\fiverm
   \textfont1=\teni\scriptfont1=\seveni\scriptscriptfont1=\fivei
   \textfont2=\tensy\scriptfont2=\sevensy\scriptscriptfont2=\fivesy
   \textfont3=\tenex\scriptfont3=\tenex\scriptscriptfont3=\tenex
   \textfont\itfam=\tenit\def\it{\fam\itfam\tenit}%
   \textfont\slfam=\tensl\def\sl{\fam\slfam\tensl}%
   \textfont\ttfam=\tentt\def\tt{\fam\ttfam\tentt}%
   \textfont\bffam=\tenbf
   \scriptfont\bffam=\sevenbf
   \scriptscriptfont\bffam=\fivebf\def\bf{\fam\bffam\tenbf}%
   \def\mib{%
      \tenmibfonts%
      \textfont0=\tenbf\scriptfont0=\sevenbf%
      \scriptscriptfont0=\fivebf%
      \textfont1=\tenmib\scriptfont1=\seveni%
      \scriptscriptfont1=\fivei%
      \textfont2=\tenbsy\scriptfont2=\sevensy%
      \scriptscriptfont2=\fivesy}%
   \tt\ttglue=.5emplus.25emminus.15em
   \normalbaselineskip=12pt
   \setbox\strutbox=\hbox{\vrule height 8.5pt depth 3.5pt width 0pt}%
   \normalbaselines\rm\singlespaced}%

\def\elevenpoint{\elevenfonts           
   \def\rm{\fam0\elevenrm}%
   \textfont0=\elevenrm\scriptfont0=\sevenrm\scriptscriptfont0=\fiverm
   \textfont1=\eleveni\scriptfont1=\seveni\scriptscriptfont1=\fivei
   \textfont2=\elevensy\scriptfont2=\sevensy\scriptscriptfont2=\fivesy
   \textfont3=\elevenex\scriptfont3=\elevenex\scriptscriptfont3=\elevenex
   \textfont\itfam=\elevenit\def\it{\fam\itfam\elevenit}%
   \textfont\slfam=\elevensl\def\sl{\fam\slfam\elevensl}%
   \textfont\ttfam=\eleventt\def\tt{\fam\ttfam\eleventt}%
   \textfont\bffam=\elevenbf
   \scriptfont\bffam=\sevenbf
   \scriptscriptfont\bffam=\fivebf\def\bf{\fam\bffam\elevenbf}%
   \def\mib{%
      \elevenmibfonts%
      \textfont0=\elevenbf\scriptfont0=\sevenbf%
      \scriptscriptfont0=\fivebf%
      \textfont1=\elevenmib\scriptfont1=\seveni%
      \scriptscriptfont1=\fivei%
      \textfont2=\elevenbsy\scriptfont2=\sevensy%
      \scriptscriptfont2=\fivesy}%
   \tt\ttglue=.5emplus.25emminus.15em
   \normalbaselineskip=13pt
   \setbox\strutbox=\hbox{\vrule height 9pt depth 4pt width 0pt}%
   \normalbaselines\rm\singlespaced}%

\def\twelvepoint{\twelvefonts\ninefonts 
   \def\rm{\fam0\twelverm}%
   \textfont0=\twelverm\scriptfont0=\ninerm\scriptscriptfont0=\sevenrm
   \textfont1=\twelvei\scriptfont1=\ninei\scriptscriptfont1=\seveni
   \textfont2=\twelvesy\scriptfont2=\ninesy\scriptscriptfont2=\sevensy
   \textfont3=\twelveex\scriptfont3=\twelveex\scriptscriptfont3=\twelveex
   \textfont\itfam=\twelveit\def\it{\fam\itfam\twelveit}%
   \textfont\slfam=\twelvesl\def\sl{\fam\slfam\twelvesl}%
   \textfont\ttfam=\twelvett\def\tt{\fam\ttfam\twelvett}%
   \textfont\bffam=\twelvebf
   \scriptfont\bffam=\ninebf
   \scriptscriptfont\bffam=\sevenbf\def\bf{\fam\bffam\twelvebf}%
   \def\mib{%
      \twelvemibfonts\tenmibfonts%
      \textfont0=\twelvebf\scriptfont0=\ninebf%
      \scriptscriptfont0=\sevenbf%
      \textfont1=\twelvemib\scriptfont1=\ninei%
      \scriptscriptfont1=\seveni%
      \textfont2=\twelvebsy\scriptfont2=\ninesy%
      \scriptscriptfont2=\sevensy}%
   \tt\ttglue=.5emplus.25emminus.15em
   \normalbaselineskip=14pt
   \setbox\strutbox=\hbox{\vrule height 10pt depth 4pt width 0pt}%
   \normalbaselines\rm\singlespaced}%

\def\fourteenpoint{\fourteenfonts\twelvefonts 
   \def\rm{\fam0\fourteenrm}%
   \textfont0=\fourteenrm\scriptfont0=\twelverm\scriptscriptfont0=\tenrm
   \textfont1=\fourteeni\scriptfont1=\twelvei\scriptscriptfont1=\teni
   \textfont2=\fourteensy\scriptfont2=\twelvesy\scriptscriptfont2=\tensy
   \textfont3=\fourteenex\scriptfont3=\fourteenex
      \scriptscriptfont3=\fourteenex
   \textfont\itfam=\fourteenit\def\it{\fam\itfam\fourteenit}%
   \textfont\slfam=\fourteensl\def\sl{\fam\slfam\fourteensl}%
   \textfont\bffam=\fourteenbf
   \scriptfont\bffam=\twelvebf
   \scriptscriptfont\bffam=\tenbf\def\bf{\fam\bffam\fourteenbf}%
   \def\mib{%
      \fourteenmibfonts\twelvemibfonts\tenmibfonts%
      \textfont0=\fourteenbf\scriptfont0=\twelvebf%
      \scriptscriptfont0=\tenbf%
      \textfont1=\fourteenmib\scriptfont1=\twelvemib%
      \scriptscriptfont1=\tenmib%
      \textfont2=\fourteenbsy\scriptfont2=\tenbsy%
      \scriptscriptfont2=\tenbsy}%
   \normalbaselineskip=17pt
   \setbox\strutbox=\hbox{\vrule height 12pt depth 5pt width 0pt}%
   \normalbaselines\rm\singlespaced}%
%
%

\def\singlespaced{
   \baselineskip=\normalbaselineskip}           


%
%
\twelvepoint
%
%
\def\begintitle{\begingroup%
\obeylines\fourteenpoint\bf\parindent=0.29truein}
\def\endtitle{\vglue 1truecm\endgroup}

\def\showheadline#1#2{\headline={\ifnum\pageno>1{\ifodd\pageno{\hfil\tenpoint #1\hfil} %
\else{\hfil\tenpoint #2\hfil}\fi} \else{\hfil}\fi}}

\def\beginauthor{\begingroup%
\obeylines\fourteenpoint\parindent=0.29truein}
\def\endauthor{\endgroup}

\def\address#1{\hbox to \hsize{\hglue 0.29in\relax
\vbox{\hsize=4.70in\relax\rightskip=0pt plus 1in\relax\noindent#1}\hfil}}

\long\def\beginaddress#1\endaddress{\vglue 6pt\address{#1}\vglue 24pt}

\def\beginabstract{\begingroup\leftskip=0.29in%
\tenpoint\noindent{\bf Abstract\ \ \ }}
\def\endabstract{\vskip 1pt minus1pt\endgroup}

\def\finalversion{\headline{\hfil}}

\def\section#1{\vskip 24pt plus4pt minus4pt\goodbreak\leftline{\bf #1}%
\vglue 12pt\nobreak\noindent\kern -0.0em}

\def\subsection#1{\vskip 12pt plus4pt minus4pt\goodbreak\leftline{\bf #1}%
\nobreak\noindent\kern -0.0em}

\def\subsubsection#1{\vskip 12pt plus4pt minus4pt\goodbreak\leftline{\it #1}%
\nobreak\noindent\kern -0.0em}

\def\begincaption#1{\begingroup\tenpoint\noindent#1\ \ \ }
\def\endcaption{\endgroup}

\newbox\@capbox                                 
\newcount\@caplines                             

\def\references{\section{REFERENCES}\tenpoint\parindent=0pt
\raggedright\rightskip=0pt plus 5em}

\def\ref#1#2{\hbox to \hsize{\vbox{\tenpoint\hsize=0.2in\relax #1\hfil}
\hfil\vtop{\hsize=4.75in\relax\tenpoint #2}}}
%
%
\vglue 1.0truein